\documentclass[a4paper]{jpconf}
\usepackage{graphicx}
\begin{document}
\title{Effect of annealing on the specific heat of optimally doped Ba(Fe$_{0.92}$Co$_{0.08}$)$_{2}$As$_{2}$}

\author{K. Gofryk$^{1}$, A. S. Sefat$^{2}$, M. A. McGuire$^{2}$, B. C. Sales$^{2}$, D.~Mandrus$^{2}$, T. Imai$^{3}$, J. D. Thompson$^{1}$, E. D. Bauer$^{1}$, F. Ronning$^{1}$}

\address{$^{1}$Condensed Matter and Thermal Physics, Los Alamos National Laboratory, Los Alamos, New Mexico 87545, USA\\
$^{2}$Materials Science and Technology Division, Oak Ridge National Laboratory, Oak Ridge, Tennessee 37831, USA\\
$^{3}$Department of Physics and Astronomy, McMaster University, Hamilton, Ontario L8S4M1, Canada}

\ead{gofryk@lanl.gov}

\begin{abstract}
We report the temperature dependence of the low-temperature specific heat down to 400~mK of the electron-doped Ba(Fe$_{0.92}$Co$_{0.08}$)$_{2}$As$_{2}$ superconductors. We have measured two samples extracted from the same batch: first sample has been measured just after preparation with no additional heat treatment. The sample shows $T_{c}$=20~K, residual specific heat $\gamma_{0}$=3.6~mJ/mol~K$^{2}$ and a Schottky-like contribution at low temperatures. A second sample has been annealed at 800~$^{o}C$ for two weeks and shows $T_{c}$=25~K and $\gamma_{0}$=1.4~mJ/mol~K$^{2}$. By subtracting the lattice specific heat, from pure BaFe$_{2}$As$_{2}$, the temperature dependence of the electronic specific heat has been obtained and studied. For both samples the temperature dependence of $C_{el}(T)$ clearly indicate the presence of low-energy excitations in the system. Their specific heat data cannot be described by single clean $s$- or $d$-wave models and the data requires an anisotropic gap scenario which may or may not have nodes.
\end{abstract}

\section{Introduction}
The discovery of superconductivity in FeAs-based materials\cite{Kamihara,chen} has re-opened new interest in superconductivity research. Unfortunately, despite a large theoretical\cite{th} and experimental\cite{ex} effort
the nature of the superconductivity in these materials including the
pairing mechanism and the symmetry of the
order parameter remains unclear.

One way to address the nature of the superconductivity is to study the low-temperature heat capacity. The measurements of the temperature dependence of the specific heat and its magnetic filed response in superconducting state give an important information about the symmetry of the order parameter\cite{njp,kg,hardy,mu3,kim1,pop}. For optimally Co doped material\cite{122b} such a analysis suggest the presence of an anisotropic gap structure which may or may not have nodes (see Ref.\cite{njp,kg,hardy,mu3}). Moreover, the specific heat displays significant residual specific heat possibly coming from the non-superconducting fraction that is present in the system\cite{kg}. Thus, the heat capacity at very low temperatures is very sensitive on the  quality of the crystals studied and it is very often dominated by a Schottky-like contribution\cite{kim}. This could make the the gap structure analysis quite complex.

In this paper, we present results of our specific heat studies of optimally doped Ba(Fe$_{0.92}$Co$_{0.08}$)$_{2}$As$_{2}$ samples. We have measured two samples extracted from the same batch: sample one has been measured just after growing and without any additional heat treatment. The sample shows $T_{c}$=20~K, residual specific heat $\gamma_{0}$=3.6~mJ/mol~K$^{2}$ and a Schottky-like contribution at low temperatures. Sample two has been annealed at 800~$^{o}C$ for two weeks with $T_{c}$=25~K and $\gamma_{0}$=1.4~mJ/mol~K$^{2}$. For this material no Schottky term is observed (see Fig.1 inset). Then using similar approach as previously reported in Ref.\cite{njp,kg} we extract and study the full electronic $T$-dependence for these two samples.

\section{Experimental details}
Single crystals were grown out of FeAs flux with a typical size of about 2$\times$1.5$\times$0.2 mm$^{3}$\cite{122b}. They crystalize in well-formed plates with the [001] direction perpendicular to the plane of the crystals. The doping level was determined by microprobe analysis. After synthesis a second piece of the sample was annealed for two weeks at 800~$^{o}C$ while the first one has not been heat treated. The heat capacity was measured down to 400~mK using a thermal relaxation method implemented in a Quantum Design PPMS-9 device.

\section{Results and discussion}

\begin{figure}[b!]
\begin{centering}
\includegraphics[width=0.58\textwidth]{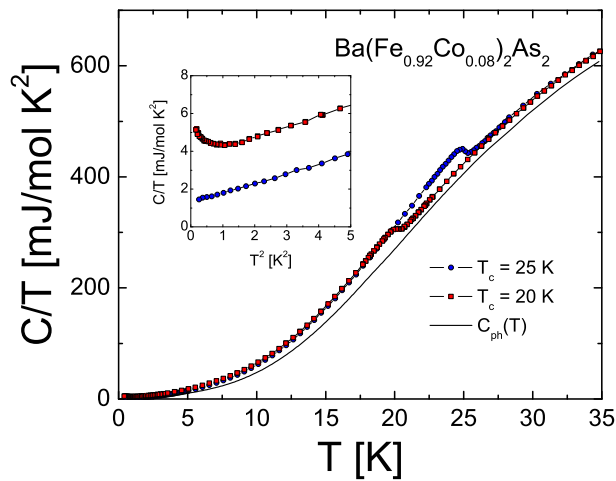}
\caption{(Color online) The temperature dependence of the heat capacity of Ba(Fe$_{0.92}$Co$_{0.08}$)$_{2}$As$_{2}$ samples. Circles: annealed sample with $T_{c}$~=~25~K. Squares: non-annealed sample with $T_{c}$~=~20~K. The solid line describes the normal state specific heat (see text). Inset: low-temperature specific heat of these two materials.}\label{fig1}
\end{centering}
\end{figure}

Fig.1 shows the temperature dependence of the specific heat of the annealed and non-annealed Ba(Fe$_{0.92}$Co$_{0.08}$)$_{2}$As$_{2}$ samples. As may be seen the annealing process has increased $T_{c}$ quite remarkably from 20~K for the non-annealed sample up to 25~K for the annealed one. It is worth noting the transition temperature of about 25~K is one of the highest reported within the Co doped Ba122 family. Above $T_{c}$ the $C_{p}(T)$ curves indicate very similar temperature dependence of the phonon part in these two materials as expected. As may be seen from the inset of Fig.1 the annealing strongly suppress the residual specific heat $\gamma_{0}$ from 3.6 to 1.4~mJ/mol~K$^{2}$ respectively for the non-annealed and annealed sample. Moreover, no Schottky-like anomaly is observed for the annealed sample. The considerably increase of $T_{c}$, decrease of $\gamma_{0}$ and lack of Schottky-type contribution at low temperature indicate significant improvement of the quality of the single crystals after the heat treatment.

In order to obtain the electronic part of the specific heat, we have used the same approach used previously for the optimal and doped  Ba(Fe$_{1-x}$Co$_{x}$)$_{2}$As$_{2}$ samples (see Ref.\cite{njp,kg}). We have assumed that the phonon contribution of the specific heat is independent of doping and we use the lattice specific heat obtained from the parent compound. We separate the lattice contribution to the specific heat of the pure Ba122 compound as $C_{ph}$~=~$C^{BaFe_{2}As_{2}}$ - $\gamma_{el}^{BaFe_{2}As_{2}}T$ where $\gamma_{el}^{BaFe_{2}As_{2}}$ is the T$\rightarrow$0 intercept of $C/T$ of BaFe$_{2}$As$_{2}$ as shown by a solid line in Fig.1. So, the electronic specific heat is simply determined by $C_{el}(T)^{Ba(Fe_{1-x}Co_{x})_{2}As_{2}}~=~C_{p}(T)^{Ba(Fe_{1-x}Co_{x})_{2}As_{2}}-C(T)_{ph}$. The obtained normal state $\gamma(T)$ from above $T_{c}$ provide perfect entropy balance for the electronic specific heat in the superconducting state below $T_{c}$ for both, non-annealed\cite{njp} and annealed samples.

\begin{figure}[b!]
\begin{centering}
\includegraphics[width=0.58\textwidth]{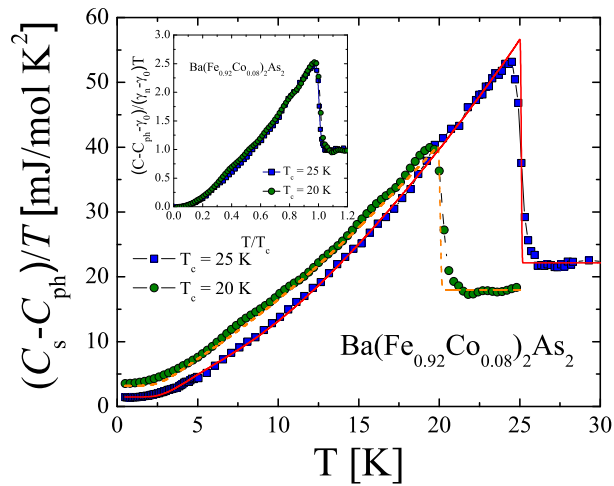}
\caption{(Color online) The temperature dependence of the non-lattice part of the specific heat of Ba(Fe$_{0.92}$Co$_{0.08}$)$_{2}$As$_{2}$ samples. Solid and dashed lines are fits of the BCS $\alpha$-model (see text). Inset: normalized temperature dependence of the superconducting part of the specific heat.}\label{fig2ab}
\end{centering}
\end{figure}

Fig.2 shows the electronic part of the specific heat of Ba(Fe$_{1-x}$Co$_{x}$)$_{2}$As$_{2}$ samples. For the non-annealed sample a small Schottky contribution below 1~K has also been subtracted. For both samples pronounced peaks are observed at the transition temperature. For the annealed sample the specific heat exhibits a jump $\Delta C/T_{c}$~=~33.6~mJ/mol~K$^{2}$ at $T_{c}$. Taking $\gamma_{n}$~=~22~mJ/mol~K$^{2}$ the ratio $\Delta C/T_{c}\gamma$~=~1.53 is slightly larger then the weak-coupling BCS value 1.43. It is also larger then $\Delta C/T_{c}\gamma$~=~1.34 obtained for the non-annealed sample\cite{njp}.

Below $T_{c}$
the $C_{el}/T$ data in Fig.2 were analyzed by the BCS $\alpha$-model:

\begin{equation}
C_{BCS}=t\frac{d}{dt}\int_{0}^{\infty}dy\left(-\frac{6\gamma\Delta_{0}}{k_{B}\pi}\right)\left[flnf+\left(1-f\right)ln\left(1-f\right)\right]
\end{equation}

where $t~=~\frac{T}{T_{c}}$, $f$ is the Fermi function $f~=~\frac{1}{e^{\frac{E}{k_{B}T}}+1}$, $E~=~\sqrt{\epsilon^{2}+\Delta^{2}}$ and $y~=~\frac{\epsilon}{\Delta}$ (see Ref.\cite{tinkham}) plus residual normal state component.

As has been shown before a single $s$-wave gap scenario cannot describe the specific heat data in Co-doped Ba122 material\cite{njp,hardy,kg,mu3}. It clearly indicates the presence of an anisotropic gap structure in this material. Moreover, the specific heat data of optimally doped Ba122 cannot be accounted by using a clean d-wave gap aproach either\cite{njp,hardy,mu3}. In order to sufficiently describe the specific heat data of annealed Ba(Fe$_{0.92}$Co$_{0.08}$)$_{2}$As$_{2}$ two separate $s$-wave gaps (solid line) have been used as shown in Fig.2. This parametrization is not unique but provides a reasonable estimate of the dominant energy scales in the gap structure. It gives parameters: $\Delta_{1}$~=~1.83~meV, $\Delta_{2}$~=~4.76~meV and $A$~=~0.72. The values of the gap are larger then the ones observed in non-annealed materials\cite{njp,hardy} $\Delta_{1}$~=~1.65~meV, $\Delta_{2}$~=~3.75~meV and $A$~=~0.62 (see dashed line in Fig.2). It is consistent with the lower transition temperature observed in the non-annealed materials. Inset of Fig.2 presents the electronic part of the specific heat of the superconducting portion of the Ba(Fe$_{0.92}$Co$_{0.08}$)$_{2}$As$_{2}$ samples. It has been obtained by subtracting the residual linear term $\gamma_{0}$, together with a small Schottky contribution below 1~K for the non-annealed sample, and normalized by ($\gamma_{n}$-$\gamma_{0}$). Interestingly, these two curves match, indicating no significant change of the gap structure in these two samples.

\section{Summary and conclusions}

In summary, using the low-temperature specific heat we explore details of the superconducting state of Ba(Fe$_{0.92}$Co$_{0.08}$)$_{2}$As$_{2}$ with $T_{c}$~=~25 and 20~K. By subtracting the phonon specific heat the temperature dependence of the electronic specific heat has been studied. The temperature variation of $C_{el}$ as well as its field dependence (not shown) cannot be described by a single isotropic $s$-wave gap, indicating the presence of an anisotropic gap structure in the system. The temperature variation of the electronic specific heat below $T_{c}$ may be well described by the presence of two superconducting gaps, pointing to a complex gap structure in the system. It has been shown recently that for optimally Co-doped BaFe$_{2}$As$_{2}$ samples a minimum of two superconducting gaps are necessary to describe the temperature dependence of the electronic specific heat\cite{njp,hardy,kg,mu3}. Moreover, for these two optimally doped samples with different quality the details of the superconducting state do not change significantly. It is in agreement with the lack of doping dependence in Ba(Fe$_{1-x}$Co$_{x}$)$_{2}$As$_{2}$\cite{kg}, indicating that the gap structure does not change significantly as a function of doping\cite{kg}. This does not address the possibility that accidental nodes may or may not to be present and could be sample, doping, and/or material dependent.\\

Work at Los Alamos National Laboratory was
performed under the auspices of the U.S. Department of
Energy, Office of Science and supported in part by the Los Alamos LDRD program. Research at Oak Ridge National Laboratory is sponsored by the Division of Material Sciences and Engineering Office of Basic Energy Sciences.

\end{document}